\documentclass[12pt,reprint,aps,groupedaddress,showpacs,nofootinbib,prx,twocolumn,superscriptaddress]{revtex4-2}
\usepackage{makecell}
\usepackage{graphicx}
\usepackage{dcolumn}
\usepackage{bm}
\usepackage{multirow}
\usepackage{mathrsfs,amsmath}
\usepackage{orcidlink}

\usepackage{array}

\begin{document}

\preprint{APS/123-QED}

\title{Accelerating TTL noise post-processing via combined coefficients and alternative TDI configuration}

\author{Xue Wang}
\email[Contact author: ]{wangxue@ucas.ac.cn}

\affiliation{Key Laboratory of Gravitational Wave Precision Measurement of Zhejiang Province, Hangzhou Institute for Advanced Study, Hangzhou, 310024, China.}
\affiliation{University of Chinese Academy of Sciences, Beijing, 100049, China.}

\author{Gang Wang\,\orcidlink{0000-0002-9668-8772}}
\email[Corresponding author:]{gwanggw@gmail.com, gwang@nbu.edu.cn}
\affiliation{Institute of Fundamental  Physics and Quantum Technology, Ningbo University, Ningbo, 315211, China}
\affiliation{Department of Physics, School of Physical Science and Technology, Ningbo University, Ningbo, 315211, China}
\affiliation{Shanghai Astronomical Observatory, Chinese Academy of Sciences, Shanghai, 200030, China}

\date{\today}

\begin{abstract}

Tilt-to-length (TTL) noise induced by angular jitter of spacecraft and test masses can affect the sensitivity of space-based gravitational-wave detectors such as LISA, Taiji, and TianQin. Such angular jitter can be measured using the differential wavefront sensing technique, enabling the modeling and subtraction of TTL noise from the data. However, owing to the multiple degrees of freedom of the detector constellation, a linear TTL model requires at least 24 parameters, while a higher-fidelity quadratic model involves up to 60 coefficients, rendering parameter estimation computationally expensive.
To accelerate parameter determination, we propose a modified parameter set obtained via a linear transformation of the original angular coupling coefficients, which effectively reduces correlations among TTL noise components. In addition, we perform parameter fitting using an alternative second-generation time-delay interferometry configuration, PD4L, rather than the fiducial Michelson configuration. These two improvements enhance the convergence speed of the fitting procedure by a factor of approximately 10 for the linear model and approximately 18 for the quadratic model. The proposed approach can therefore substantially improve the efficiency of TTL noise calibration in space-based gravitational-wave detectors.

\end{abstract}

\maketitle

\section{Introduction}

Space-based gravitational wave (GW) detectors are designed to probe the low-frequency band from $0.1~\mathrm{mHz}$ to $1~\mathrm{Hz}$, bridging the observational gap between ground-based interferometers operating above $\sim 1~\mathrm{Hz}$ and pulsar timing arrays sensitive to nanohertz frequencies. Major space-based missions under development include the Laser Interferometer Space Antenna (LISA) \cite{amaroseoane2017laser,Colpi:2024xhw}, Taiji \cite{Hu:2017mde,10.1093/ptep/ptaa083}, and TianQin \cite{TianQin:2015yph,10.1093/ptep/ptaa114}.
Achieving the target sensitivity of space-based interferometers requires displacement measurements with precision at the picometer level. Such performance demands stringent suppression and calibration of multiple noise sources. Laser frequency noise, which exceeds targeting sensitivity by several orders of magnitude, is mitigated through time-delay interferometry (TDI) technology \cite[and references therein]{1997SPIE.3116..105N,1999ApJ...527..814A,Tinto:2020fcc}. In addition to laser frequency noise, tilt-to-length (TTL) noise constitutes a significant limitation at high frequencies. TTL noise arises from the coupling between angular jitter within the optical system and the effective optical path length, thereby contaminating the interferometric phase measurements.

TTL noise involves multiple coupling models and mechanisms. The models include the laser source model, the mirror model, and the receiver system model, while key mechanisms involve piston noise, lever noise \cite{Hartig_2022,Hartig_2023}, and phase errors induced by optical aberrations \cite{Sasso_2018_far-field,Kenny_2021,Weaver_2022,Xiao_2023}. Extensive research has been conducted on TTL noise suppression using imaging systems \cite{Chwalla_2016,Tröbs_2018,PhysRevApplied.14.014030,TTLtwo-lens}. Through precise optical alignment and adjustment, the TTL coupling can be effectively reduced. However, this approach alone remains insufficient for space-based GW detection \cite{PhysRevD.110.022003}. Therefore, TTL noise post-processing serves as a further essential mitigation approach.

In addition to measuring relative displacement, space-based interferometers employ Differential Wavefront Sensing (DWS) \cite{Hechenblaikner:10} to monitor angular jitter of the Movable Optical System Subassembly (MOSA) \cite{amaroseoane2017laser,GAO2023107287,Paczkowski:2022nrt}. Since MOSA jitter constitutes the dominant contributor to TTL noise, post-processing approaches typically model TTL noise as a function of the measured angular motion and a set of coupling coefficients, which are subsequently subtracted from the interferometric data. Two main strategies exist for determining these coupling coefficients. One relies on dedicated calibration maneuvers that intentionally excite angular motion \cite{spacecraft_maneuver_design}, while the other estimates the coefficients directly from science data through long-term correlations between DWS signals and interferometric measurements. In both cases, reliable estimation of TTL coupling coefficients must be performed after TDI processing, as laser frequency noise dominates the raw measurements prior to TDI.

The propagation of TTL noise within second-generation TDI combinations has been investigated in detail in Refs.~\cite{PhysRevD.110.022003,Houba:2022wni}. Numerical demonstrations of TTL noise post-processing and subtraction were subsequently carried out in Refs.~\cite{Paczkowski:2022nrt,George:2022pky}, confirming its effectiveness in suppressing residual TTL noise. Extensions of these studies have considered alternative TDI schemes and algorithmic frameworks, including the application of TDI Infinity \cite{houba2023time}. Most existing analyses, however, have focused on linear TTL coupling models. An exception is Ref.~\cite{wang2024post}, which implemented second-order TTL models and examined the physical origin of residual TTL noise, albeit without incorporating additional secondary noise sources.

In this work, we extend previous studies by incorporating laser frequency noise, acceleration noise, and optical metrology system (OMS) noise into comprehensive numerical simulations, thereby enabling a more realistic assessment of TTL noise modeling and subtraction. We first estimate TTL coupling coefficients using the second-generation TDI configuration PD4L, which has been shown to exhibit improved robustness relative to the fiducial Michelson combination \cite{Wang:2011,Wang:2025mee,Wang:2025voa}. By analyzing the posterior distributions of the inferred coefficients, we identify strong parameter degeneracies and introduce a transformed parameter set obtained through a linear combination of the original coefficients to reduce mutual coupling.
Based on this transformed parameterization, we adopt the deterministic fitting algorithm \textsf{Minuit} \cite{James:1975dr,iminuit} to estimate TTL coupling coefficients, avoiding the high computational cost associated with Markov Chain Monte Carlo and nested sampling methods. To quantitatively assess the efficiency gains enabled by this approach, we perform parameter fitting using two TDI configurations (Michelson and PD4L) and two TTL models (linear and quadratic). We show that, for the Michelson configuration with a linear model, the efficiency improvement remains limited due to strong parameter degeneracy. In contrast, substantial acceleration is achieved when either higher-order models or alternative TDI configurations are employed. In particular, the combination of the transformed parameter set and the PD4L configuration yields speed-ups of up to a factor of $\sim 18$ for quadratic TTL models.

This paper is organized as follows. In Section~\ref{sec:ttl_noise}, we analyze the physical mechanisms underlying TTL noise and justify the use of MOSA jitter in both noise generation and post-processing. Section~\ref{sec:ttl_noise_simulation} describes the numerical framework used to simulate TTL noise and other instrumental noises. In Section~\ref{sec:post_processing}, we present the parameter transformation, investigate coefficient degeneracies, and demonstrate the resulting improvements in fitting efficiency.

\section{TTL noise} \label{sec:ttl_noise}

\begin{figure*}[hbt]
\centering
\includegraphics[scale=0.4]{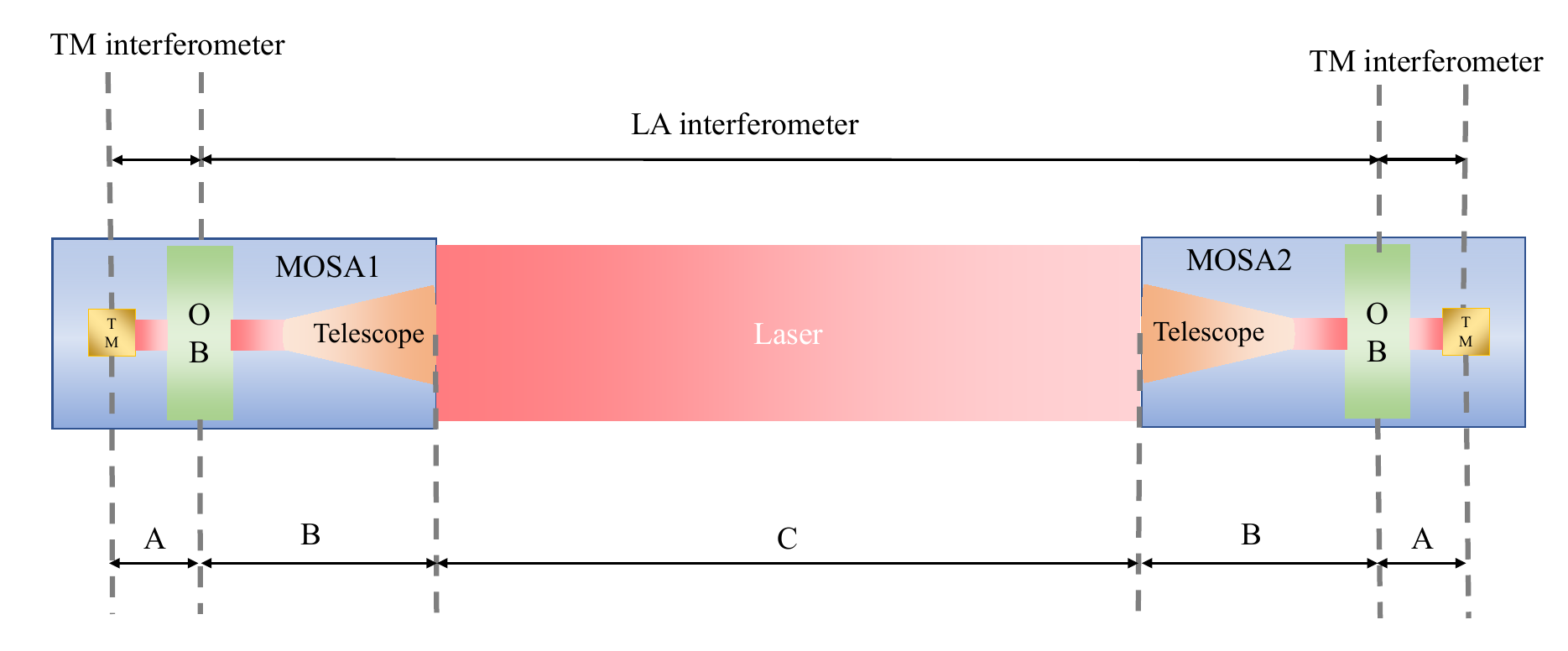}
\caption{A: TM interferometer TTL noise; B: LA interferometer receiver TTL noise; C: LA interferometer transmitter TTL noise.}
\label{Noise_location}
\end{figure*}

\begin{figure}[hbt]
\centering
\includegraphics[scale=0.5]{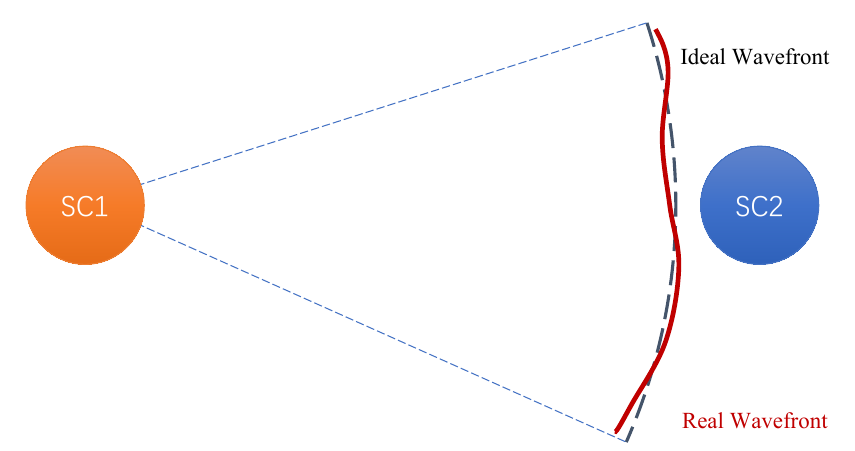}
\caption{TTL noise due to far-field phase error.}
\label{LA_interferometer_transmitter_TTL_noise}
\end{figure}

The TTL noise in one arm is depicted in Fig. \ref{Noise_location}. The measurement for each interferometric arm is performed by two MOSAs. Each MOSA comprises a telescope, an optical bench (OB), an gravitational reference sensor (GRS), inside which a freely floating Test Mass (TM) is located. All of those are rigidly connected \cite{SC_nonlinear_dynamics_model}. The Long-Arm interferometer (LA interferometer, also known as inter-spacecraft interferometer or science interferometer) measures the relative displacement between the OBs on different spacecrafts. The TM interferometer measures the relative displacement between the OB and the TM. Based on its location, TTL noise can be classified into: TM interferometer TTL noise, LA interferometer receiver TTL noise, and LA interferometer transmitter TTL noise.
\begin{itemize}
    \item TM Interferometer TTL Noise: jitter of the MOSA will cause relative motion between the optical bench and the TM. Since the TM acts as a mirror in the interferometer, this effectively creates a jittering mirror within the TM interferometer, thereby producing TTL noise.
    \item LA Interferometer Receiver TTL Noise: when the MOSA, acting as a receiver, rotates relative to the incoming beam, the laser enters the telescope at a varying angle and subsequently enters the optical bench at a varying angle. TTL noise is generated when the laser passes through the telescope at varying angles and subsequently enters the interferometer at differing incidence angles.
    \item LA Interferometer Transmitter TTL Noise (Fig.~\ref{LA_interferometer_transmitter_TTL_noise}): when the MOSA rotates as a transmitter, it is equivalent to a rotation of the laser source. Due to aberrations in the outgoing beam, the wavefront deviates from an ideal spherical shape after long-distance propagation. A phase shift is introduced in the laser received by the telescope at the receiver end when the transmitter rotates.
\end{itemize}

Furthermore, the Point-Ahead Angle Mechanism (PAAM) is installed in the outgoing beam path. Jitter of the PAAM causes jitter of the outgoing beam, which further contributes to the LA interferometer transmitter TTL noise \cite{houba2022lisapointaheadanglecontrol,inproceedings}.
The magnitude of different TTL noise is different. By comparing the magnitudes of various TTL noises, certain couplings can be neglected, thereby simplifying the analysis and post-processing procedures.
The beam jitter induced by the PAAM is compressed by the telescope's magnification factor M after passing through the telescope. Additionally, since the PAAM is a compact piezoelectric component, its angular jitter is relatively small. Consequently, the jitter caused by the PAAM is significantly lower than that of the MOSA, allowing the TTL noise generated by the PAAM to be neglected.

Both TM interferometer TTL noise and LA interferometer receiver TTL noise can be suppressed by an imaging system \cite{Chwalla_2016,Tröbs_2018,PhysRevApplied.14.014030,TTLtwo-lens}. The angular jitter in the TM interferometer is equal to the angular jitter of the MOSA.  From the receiver's perspective, the jitter of the MOSA relative to the incident beam is equivalent to an opposite-direction jitter of the incident beam itself. At the telescope's large pupil, the jitter magnitude of the incident beam is equal to that of the MOSA relative to the incident beam. When the jittered laser passes through the telescope and enters the optical interferometer through the small pupil, the angular jitter is amplified by the telescope's magnification factor. Consequently, the LA Interferometer Receiver TTL Noise is more crucial than TM Interferometer TTL Noise.

For the same alignment precision, the same level of TTL noise suppression is attained by the imaging system . The telescope magnification factor is 134 for LISA \cite{Sankar_2020_telescope} and 100 for Taiji \cite{GAO2023107287}. Consequently, the TTL noise at the receiver of the LA interferometer is significantly greater than that of the TM interferometer. Furthermore, since the angular jitter in both the TM and LA interferometers originates from the jitter of the MOSA, the simulation and post-processing of TM interferometer TTL noise can be combined with that of the LA interferometer.

Additionally, lateral translation of the MOSA also generates TTL noise. However, this type of TTL coupling fundamentally corresponds to a displacement of the optical interferometer along its optical axis. During TDI processing, this component of TTL noise is canceled when the measurement signals from the TM and LA interferometers are combined. Therefore, the impact of MOSA lateral motion can also be neglected.

Based on the preceding analysis, although the coupling mechanisms of TTL noise are complex, the predominant source of jitter for major TTL noise components is attributed to the MOSA. Even when the coupling paths differ, provided the jitter originates from the MOSA, the corresponding TTL noise components can be consolidated. Consequently, for both noise generation and post-processing, it is sufficient to consider only the jitter of the MOSA.

\section{Noise Simulation} \label{sec:ttl_noise_simulation}

\subsection{TTL noise generation}

\begin{figure}[hbt]
\centering
\includegraphics[scale=0.5]{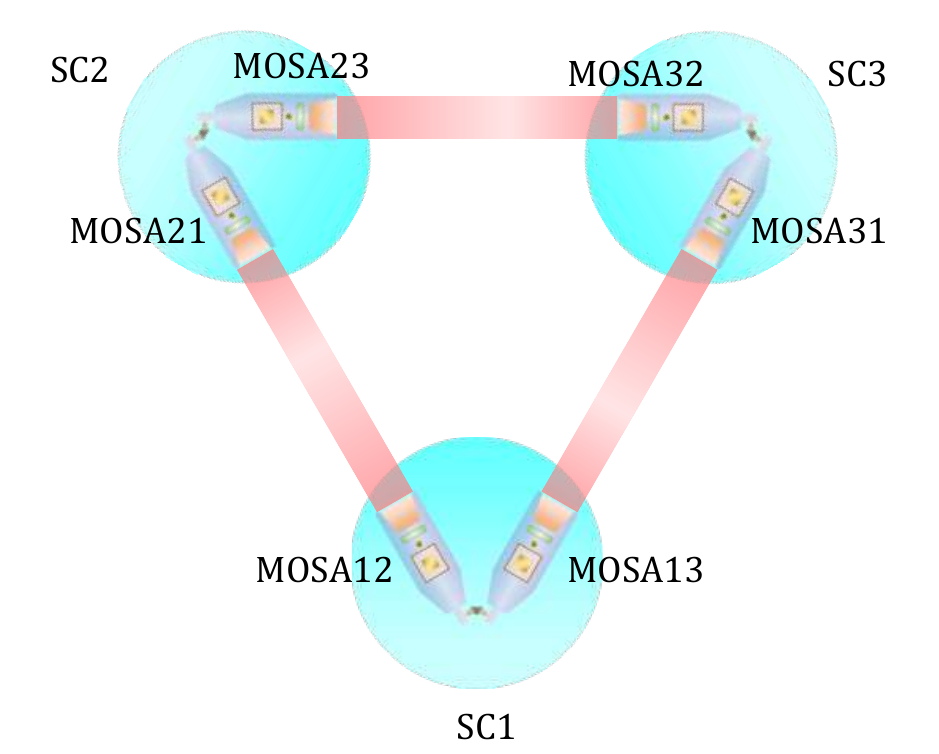}
\caption{The constellation layout and notations of detector.}
\label{fig:Gravitational_Wave_Constellation}
\end{figure}

A space-based gravitational-wave detector consists of three spacecraft forming a triangular interferometric constellation, as illustrated in Fig.~\ref{fig:Gravitational_Wave_Constellation}. The spacecraft are labeled as $\mathrm{SC}1$, $\mathrm{SC}2$, and $\mathrm{SC}3$, and the indices $i$, $j$, and $k$ are used throughout to denote the corresponding spacecraft. Each spacecraft hosts two MOSAs. We denote by MOSA$ij$ the MOSA on SC$i$ pointing toward SC$j$, and by MOSA$ik$ the MOSA pointing toward $\mathrm{SC}k$. The same convention applies to the OBs and to the interferometric measurements associated with them.

\begin{figure}[hbt]
\centering
\includegraphics[scale=0.5]{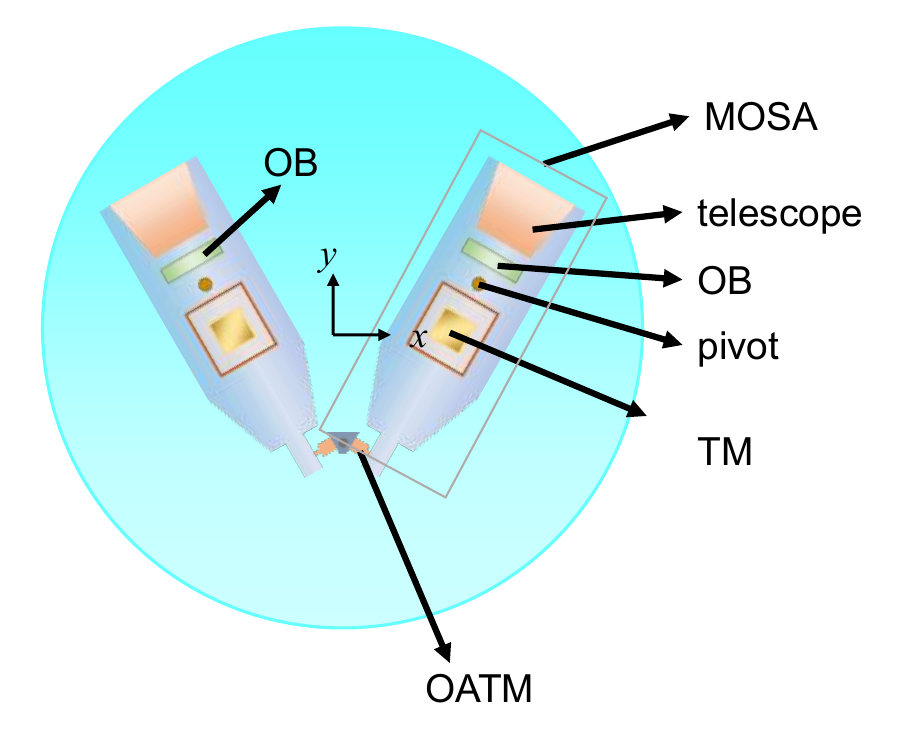}
\caption{Spacecraft coordinate system.}
\label{fig:spacecraft}
\end{figure}

Pointing jitter is defined as the angular fluctuation of a MOSA relative to the incident laser beam. It arises from two primary sources: the spacecraft attitude jitter and the relative jitter of the MOSA with respect to the spacecraft, which is driven by the optical assembly tracking mechanism (OATM). The spacecraft coordinate system is defined with its origin at the spacecraft center of mass, as shown in Fig.~\ref{fig:spacecraft}. The $x$--$y$ plane coincides with the constellation plane, the $z$-axis is normal to the plane, and the $y$-axis bisects the nominal pointing directions of the two MOSAs \cite{SC_nonlinear_dynamics_model}.

The spacecraft attitude motion is described by three rotational degrees of freedom: yaw (rotation about the $z$-axis), pitch (rotation about the $x$-axis), and roll (rotation about the $y$-axis). The OATM primarily actuates the MOSA around the yaw axis. Since the present study focuses on modeling pointing jitter from the actuator perspective, noise contributions from non-operational directions are neglected.
The jitter noise levels follow the specifications adopted in Refs.~\cite{Paczkowski:2022nrt,George:2022pky,GAO2023107287}. The amplitude spectral density of the spacecraft attitude jitter for all three rotational degrees of freedom is modeled as
\begin{equation}
S^{1/2}_{{\rm SC}(\rm yaw,pitch,roll)} =\frac{10 \ {\rm{nrad}}}{\rm\sqrt{Hz}}\sqrt{1+\left(\frac{0.8{\rm mHz}}{f}\right)^4}.
\label{equation12}
\end{equation}
while the relative yaw jitter of the MOSA with respect to the spacecraft induced by the OATM is given by
\begin{equation}
S^{1/2}_{{\rm MOSA\_SC}(\rm yaw)}=\frac{10 \ {\rm{nrad}}}{\rm\sqrt{Hz}}\sqrt{1+\left(\frac{0.8{\rm mHz}}{f}\right)^4}.
\label{equation13}
\end{equation}

Directly generating random jitter realizations from the above spectra leads to pronounced low-frequency drift. Over long simulation durations, the accumulated pointing deviation becomes unphysical. To mitigate this effect, the noise spectra are truncated below $0.1~\mathrm{mHz}$ in all numerical simulations, and frequency components below this cutoff are excluded.
Based on the spacecraft attitude jitter and the MOSA-relative jitter, the effective angular jitter of each MOSA with respect to the incident beam can be computed. For the yaw component, the relations are
\begin{align}
\mathrm{yaw}_{ij} = & \mathrm{yaw}_{\mathrm{SC}i}+ \mathrm{yaw}_{\mathrm{MOSA}ij}, \label{equation14} \\
\mathrm{yaw}_{ik} = & \mathrm{yaw}_{\mathrm{SC}i} + \mathrm{yaw}_{\mathrm{MOSA}ik}. \label{equation15}
\end{align}
For the pitch component, geometric projection yields
\begin{align}
\mathrm{pitch}_{ij} = & \cos 30^\circ \times \mathrm{pitch}_{\mathrm{SC}i}+ \sin 30^\circ \times \mathrm{roll}_{\mathrm{SC}i} \label{equation16}, \\
\mathrm{pitch}_{ik} = & \cos 30^\circ \times \mathrm{pitch}_{\mathrm{SC}i}-\sin 30^\circ \times \mathrm{roll}_{\mathrm{SC}i} \label{equation17}.
\end{align}

Under the above framework, TTL noise in the LA interferometer arises predominantly at both the transmitter and receiver. For the receiver, the TTL coupling is modeled using a second-order expansion in the pointing angles. The first-order contribution is expressed as
\begin{equation}
\mathrm{TTL}^\mathrm{Rx1}={\rm C}_{y} \mathrm{yaw}+{\rm C}_{p} \mathrm{pitch},
\label{eq:1st_order_model}
\end{equation}
while the second-order contribution is given by
\begin{equation}
\mathrm{TTL}^\mathrm{Rx2}={\rm C}_{yy} \mathrm{yaw}^2+{\rm C}_{pp} \mathrm{pitch}^2+{\rm C}_{yp} \mathrm{yaw\times pitch}.
\label{eq:2nd_order_model}
\end{equation}
The total receiver-side TTL noise is therefore
\begin{equation}
\mathrm{TTL}^\mathrm{Rx}= \mathrm{TTL}^\mathrm{Rx1}+ \mathrm{TTL}^\mathrm{Rx2}.
\label{eq:1st_2nd_order_model}
\end{equation}

In the numerical simulations, the coupling coefficients are randomly generated subject to two constraints. First, the sum of the first-order coupling coefficients is required to be smaller than $2.3~\mathrm{mm/rad}$, corresponding to a peak-to-valley TTL displacement below $920~\mathrm{nm}$ within a $\pm200~\mu\mathrm{rad}$ angular range. Second, the peak-to-valley amplitude of the second-order TTL contribution is limited to $10\%$ of that of the first-order term, i.e., $92~\mathrm{nm}$. This constraint enforces a controlled hierarchy between linear and nonlinear coupling strengths.

The transmitter-side TTL coupling is dominated by far-field wavefront distortions. This effect is modeled by first generating random wavefront aberrations using the first 15 Zernike polynomials, with a root-mean-square (RMS) amplitude of $\lambda/20$ \cite{Sankar_2020_telescope}. For an operating wavelength of $\lambda = 1064~\mathrm{nm}$, this corresponds to an RMS aberration of $53~\mathrm{nm}$. Diffraction propagation is then performed over a distance of $3\times10^{6}~\mathrm{km}$ using the algorithm described in \cite{wang2024post}, yielding the angular-to-length coupling for the transmitter.

\begin{figure*}[htb]
\includegraphics[width=0.48\textwidth]{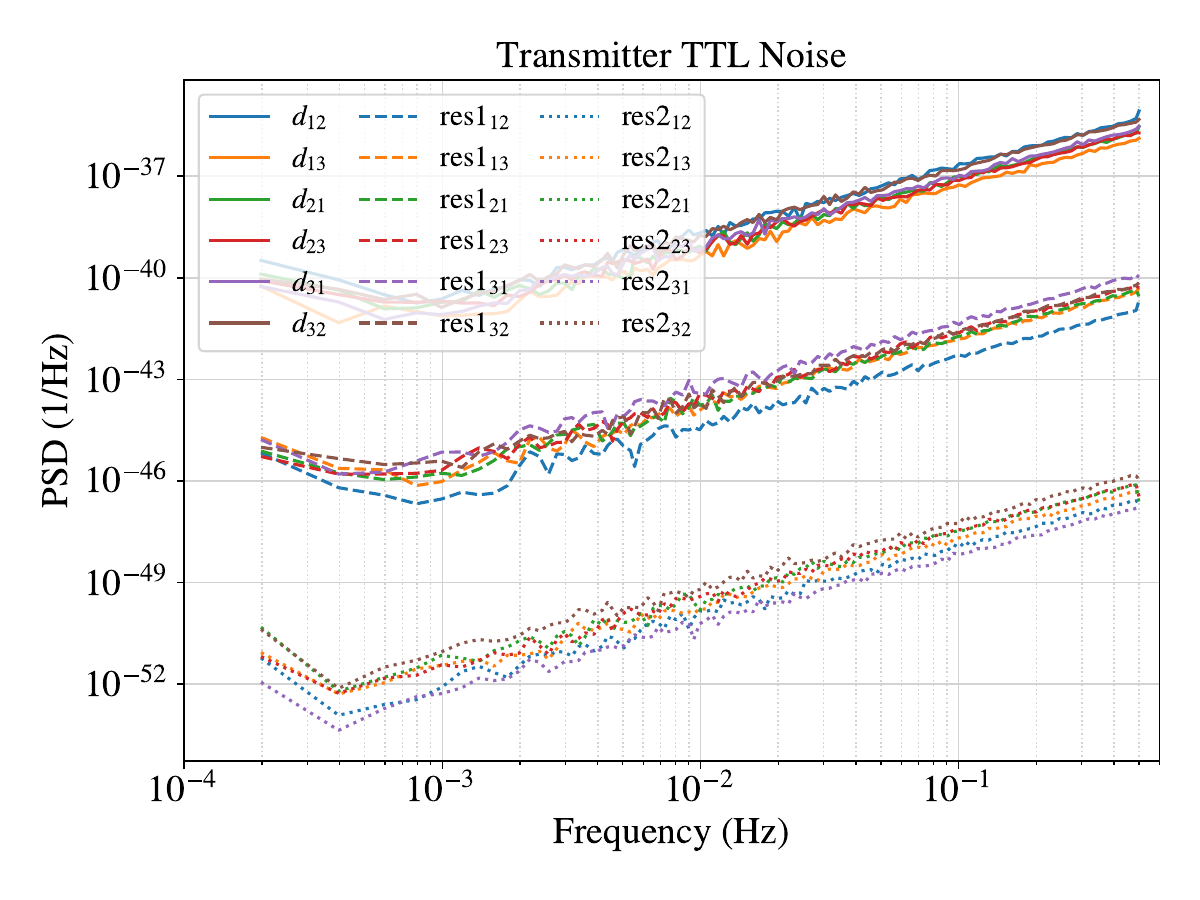}
\includegraphics[width=0.48\textwidth]{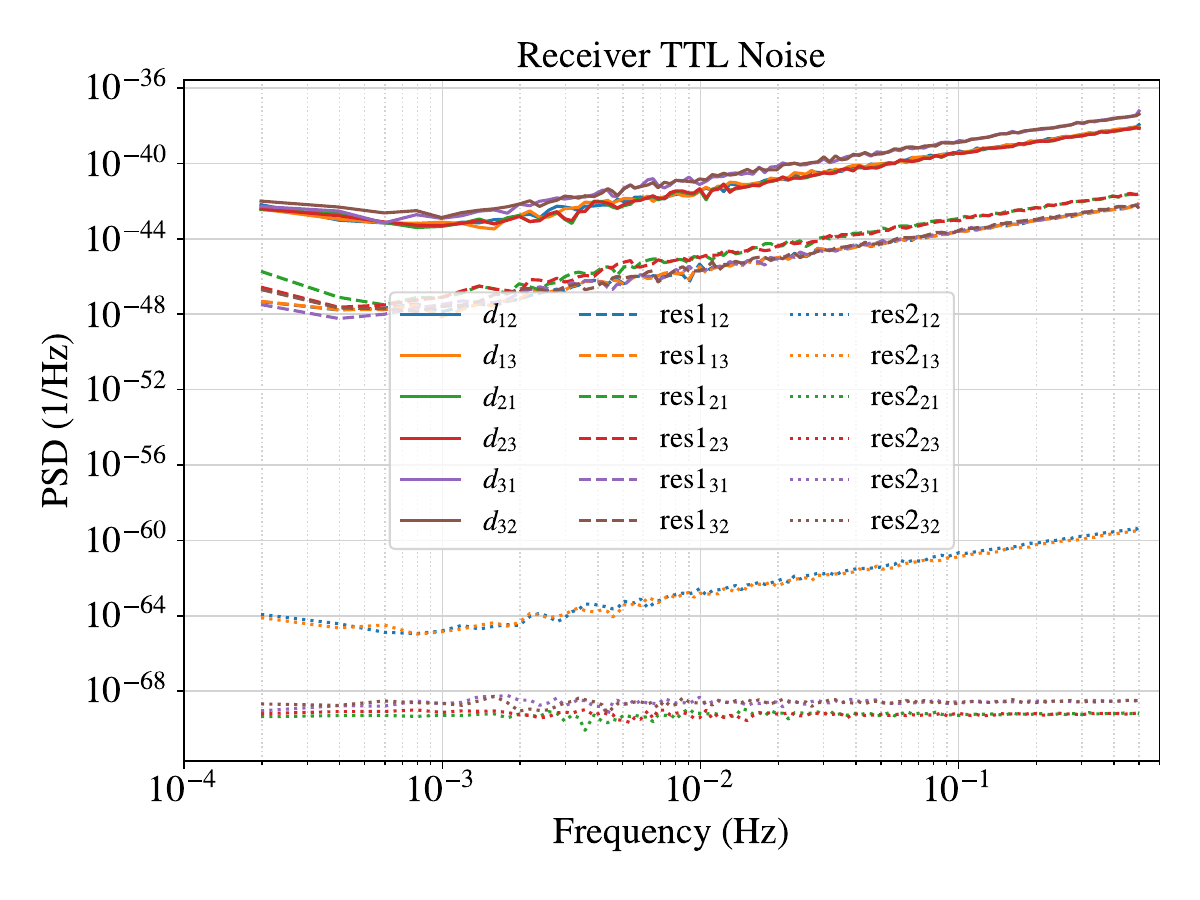}
\caption{\label{fig:TTL_noise_links} TTL noise PSDs at the transmitters (left panel) and receivers (right panel). For each panel, six solid curves ($d_{ij}$) correspond to the simulated TTL noise of the six inter-spacecraft links. The dashed (res1$_{ij}$) and dotted (res2$_{ij}$) curves represent the residuals after linear and quadratic model fitting and subtraction, respectively, applied directly to the simulated noise. This procedure provides a direct validation of the consistency between the adopted coupling models and the injected TTL noise.
}
\end{figure*}

For each MOSA, independent transmitter-side and receiver-side coupling models are generated, and the corresponding TTL noise is computed using the associated jitter realizations. As a result, a total of 12 independent TTL noise components are produced in a single simulation. All of these components must be identified and calibrated during TTL post-processing, posing a substantial challenge for data analysis.
It is worth noting that the jitter of the two MOSAs on the same spacecraft is partially correlated, since both are synthesized from the same spacecraft attitude motion combined with independent MOSA-relative jitter. This intrinsic correlation further complicates the separation and subtraction of individual TTL noise contributions.

Fig.~\ref{fig:TTL_noise_links} presents the power spectral densities (PSDs) of TTL noise at the transmitters (left panel) and receivers (right panel). In each panel, three groups of curves are shown, and six curves in each group correspond to the six inter-spacecraft links in the constellation. For each link, the solid lines represent the simulated TTL noise PSDs, serving as the reference input. The dashed lines show the residuals obtained after subtracting the fitted TTL noise using the linear (first-order) model, while the dotted lines correspond to the residuals after quadratic (second-order) model fitting and subtraction.
All fittings are performed directly on the simulated TTL noise, without the inclusion of other instrumental noises, in order to assess the intrinsic consistency between the adopted models and the injected noise. For both the transmitter and receiver cases, the linear model achieves a substantial reduction of the TTL noise across the frequency band. The quadratic model provides a further systematic suppression, yielding residual levels that are significantly lower than those of the linear fit for all six links. This comparison demonstrates that the second-order model captures additional nonlinear features of the TTL coupling and offers a markedly improved representation of both transmitter and receiver TTL noise.

\subsection{TTL noise incorporation}

The interferometric architecture adopts the baseline design of the LISA mission, as detailed in \cite{Otto:2012dk,Otto:2015}. Each OB supporting a trio of interferometers: the long-arm science interferometer ($s_{ji}$), the test-mass interferometer ($\varepsilon_{ij}$), and the reference interferometer ($\tau_{ij}$). For the receiving benches, the interferometric outputs are simulated as:
\begin{equation} \label{eq:s_epsilon_tau}
\begin{aligned}
   s_{ij} (t) = & p_{ji}(t - L_{ji}) - p_{ij}(t) + n^{\mathrm{oms}}_{ij}(t) \\  & + \mathrm{TTL}^\mathrm{Tx}_{ji} (t - L_{ji}) - \mathrm{TTL}^\mathrm{Rx}_{ij} (t) , \\
   \varepsilon_{ij} (t) = & p_{ik}(t) - p_{ij}(t) - 2 n^{\mathrm{acc}}_{ij}(t), \\
   \tau_{ij} (t) = & p_{ik}(t) - p_{ij}(t).
\end{aligned}
\end{equation}
where $p_{ij}$ denotes the laser frequency noise on the OB of SC$i$ facing SC$j$, while $n^{\mathrm{oms}}_{ij}$ and $n^{\mathrm{acc}}_{ij}$ correspond to the OMS noise and the TM acceleration noise for MOSA host on SC$i$ facing to SC$j$, respectively. $L_{ji}$ is the ranging time from SC$j$ to SC$i$. 
The noise model adopted in this study includes both acceleration noise and OMS noise, following the baseline parameters of the Taiji mission \cite{Luo:2019zal}:
\begin{equation} \label{eq:noise_budgets}
\begin{aligned}
& \sqrt{ S_{\mathrm{acc}} } = 3 \, \frac{\mathrm{fm/s^2}}{\sqrt{\mathrm{Hz}}} \sqrt{1 + \left(\frac{0.4 \, \mathrm{mHz}}{f} \right)^2 }  \sqrt{1 + \left(\frac{f}{8 \, \mathrm{mHz}} \right)^4 }, \\
& \sqrt{ S_{\mathrm{oms}} } = 8 \, \frac{\mathrm{pm}}{\sqrt{\mathrm{Hz}}} \sqrt{1 + \left(\frac{2 \, \mathrm{mHz}}{f} \right)^4 }.
 \end{aligned}
\end{equation}
TTL$^\mathrm{Tx}_{ji}$ is TTL noise generated on SC$j$ sending to SC$i$, and TTL$^\mathrm{Rx}_{ij}$ is the TTL noise caused on the receiver. The TTL noises already include the components from pitch and yaw motions. We admit that the TTL noises are also generated in the TM interferometers $\varepsilon_{ij}$ due to the tilt of TM, this is expected to be much lower than the TTL noise in science interferometers and is ignored in this investigation \cite{PhysRevD.110.022003}. Instead of correcting TTL noise after the TDI process, we model TTL noise and subtract them in the science interferometers $s_{ij}$ and obtain the corrected data as 
\begin{equation}
    s^\mathrm{cor}_{ij} (t) = s_{ij} (t) - \mathrm{TTL}^\mathrm{Tx,model}_{ji} (t - L_{ji}) - \mathrm{TTL}^\mathrm{Rx, model}_{ij} (t), 
\end{equation}
where $\mathrm{TTL}^\mathrm{Tx,model}_{ji}$ and $ \mathrm{TTL}^\mathrm{Rx, model}_{ij}$ are the linear or quadratic models for the transmitter and receiver, respectively.
These interferometric data streams are synthesized into a combined observable $\eta_{ji}$ for TDI process:
\begin{equation} \label{eq:eta}
\begin{aligned}
  \eta_{ji} &= s^\mathrm{cor}_{ij} + \frac{1}{2} \left[ \tau_{ij} - \varepsilon_{ij} + \mathcal{D}_{ji} ( 2 \tau_{ji} - \varepsilon_{ji} - \tau_{jk} ) \right] \\
  & \quad \text{for } (\mathrm{SC}2 \rightarrow \mathrm{SC}1), (\mathrm{SC}3 \rightarrow \mathrm{SC}2), \text{ and } (\mathrm{SC}1 \rightarrow \mathrm{SC}3), \\
  \eta_{ji} &= s^\mathrm{cor}_{ij} + \frac{1}{2} \left[ \tau_{ij} - \varepsilon_{ij}  + \mathcal{D}_{ji} ( \tau_{ji} - \varepsilon_{ji}    ) + \tau_{ik} - \tau_{ij} \right] \\
   & \quad \text{for } (\mathrm{SC}1 \rightarrow \mathrm{SC}2), (\mathrm{SC}2 \rightarrow \mathrm{SC}3), \text{ and } (\mathrm{SC}3 \rightarrow \mathrm{SC}1),
\end{aligned}
\end{equation}
where $D_{ji}$ is a delay operator $D_{ji} \ y (t) = y(t - L_{ji})$. Time-dependent arm lengths are extracted from a randomly selected epoch of a numerical orbit simulation \cite{Wang:2017aqq}\footnote{\url{https://github.com/gw4gw/LISA-Like-Orbit}}.

\section{Post-processing of TTL noise} \label{sec:post_processing}

\subsection{TDI configurations}

TDI is employed to suppress laser frequency noise by combining inter-spacecraft measurements to synthesize virtual equal-arm interferometry \cite[and references therein]{1997SPIE.3116..105N,1999ApJ...527..814A,Tinto:2020fcc}. By choosing different initial spacecraft and propagation sequences, three Michelson observables, conventionally denoted as X1, Y1, and Z1, can be constructed. The first of these, X1, can be written in the notation of \citet{Vallisneri:2005ji} as
\begin{equation} \label{eq:X1_measurement}
{\rm X1:} \ \overrightarrow{121313121} \ \overleftarrow{131212131}.
\end{equation}
Here, the arrows indicate the temporal ordering of the light propagation: the symbol ``$\rightarrow$'' denotes forward time evolution from left to right, while ``$\leftarrow$'' represents backward-time propagation from left to right. The numbers label the corresponding spacecraft.

\begin{figure}[htb]
\includegraphics[width=0.3\textwidth]{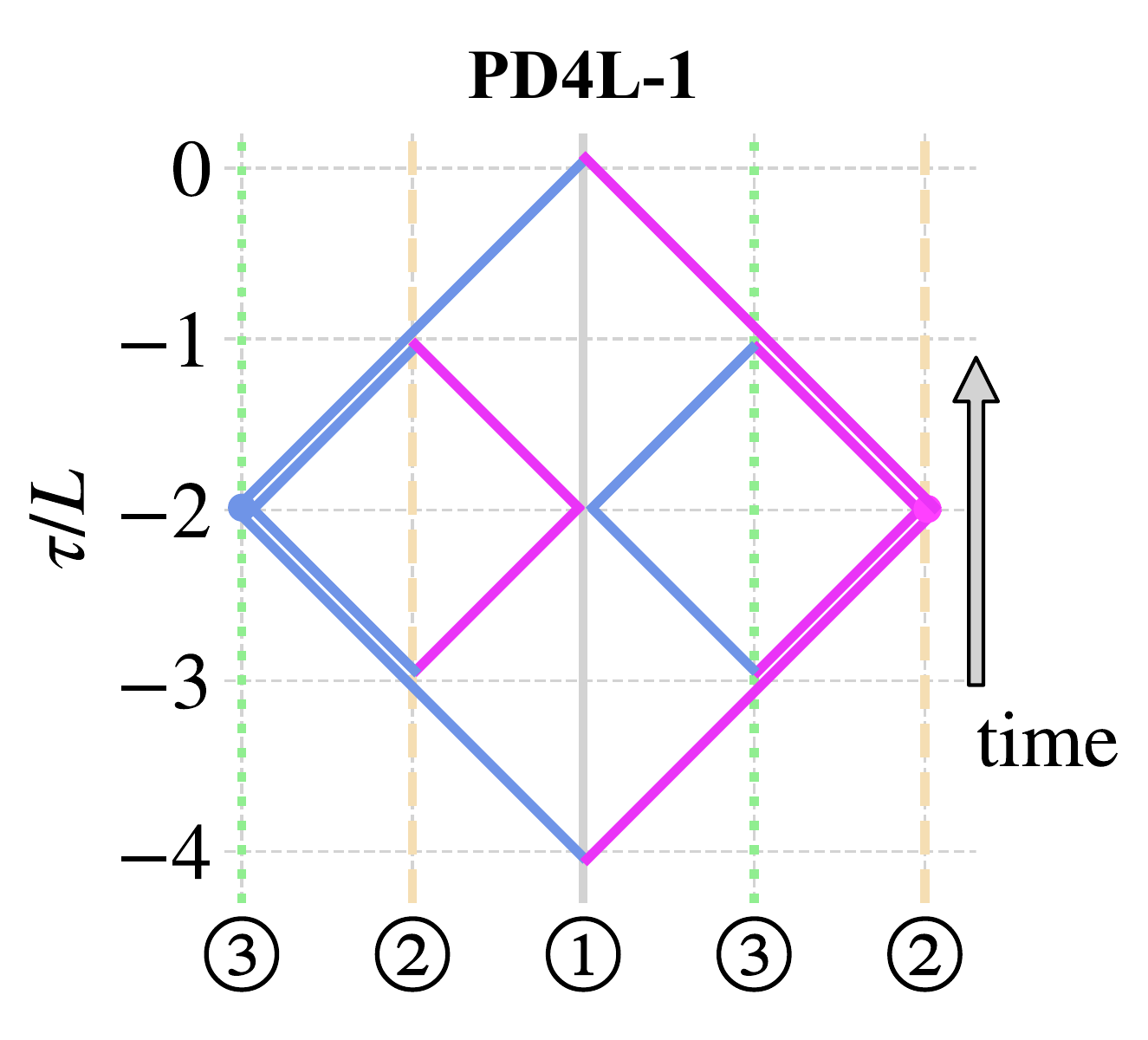}
\caption{\label{fig:PD4L_diagram} Diagram of the PD4L configuration (reused from \cite{Wang:2025mee}). The vertical lines represent the temporal evolution of the spacecraft, labeled \textcircled{$i$} for SC$i$ ($i=1,2,3$). The vertical axis indicates the time delay relative to the TDI initial epoch $\tau=0$. For visual clarity and to avoid path overlap at non-integer delays, auxiliary lines are shown for SC2 (orange dashed) and SC3 (green dotted). Blue lines illustrate the $\rightarrow$ paths in Eq.~\eqref{eq:PD4L_path}, and magenta lines indicate the $\leftarrow$ paths.
}
\end{figure}

However, the Michelson TDI combination exhibits a large number of null frequencies, which can reduce its robustness in data analysis and noise characterization \cite{Wang:2024alm,Wang:2024hgv}. In this work, we therefore adopt an alternative second-generation TDI configuration, PD4L, which was initially developed in \cite{Wang:2011} and has been shown to be more robust than the fiducial Michelson \cite{Wang:2025mee,Wang:2025voa}. The three PD4L observables are denoted as PD4L-1, PD4L-2, and PD4L-3. The first of these can be expressed as
\begin{equation} 
\textrm{PD4L-1:}\ \overrightarrow{1 2 3 2} \ \overleftarrow{2 1 2} \ \overrightarrow{2 3 2 1} \ \overleftarrow{1 3 2 3} \ \overrightarrow{3 1 3} \ \overleftarrow{3 2 3 1} \label{eq:PD4L_path},
\end{equation}
and its geometric representation is shown in Fig.~\ref{fig:PD4L_diagram}. In the diagram, the blue lines indicate the $\rightarrow$ propagation paths, while the magenta lines represent the $\leftarrow$ paths.

For a given TDI configuration, a triplet of channels $(a,b,c)$ is obtained through cyclic permutation of the spacecraft indices. The noise properties of these channels are characterized by the covariance matrix
\begin{equation} \label{eq:covmat_approximation}
\begin{bmatrix}
S_\mathrm{a} & S_\mathrm{ab} & S_\mathrm{ac} \\
S_\mathrm{ba} & S_\mathrm{b} & S_\mathrm{bc} \\
S_\mathrm{ca} & S_\mathrm{cb} & S_\mathrm{c}
\end{bmatrix}
\simeq
\begin{bmatrix}
S_\mathrm{a} & S_\mathrm{ab} & S_\mathrm{ab} \\
S_\mathrm{ab} & S_\mathrm{a} & S_\mathrm{ab} \\
S_\mathrm{ab} & S_\mathrm{ab} & S_\mathrm{a}
\end{bmatrix},
\end{equation}
where the diagonal elements $S_\mathrm{a}$ denote the PSD of channel $a$, and the off-diagonal elements $S_\mathrm{ab}$ represent the cross-spectral density (CSD) between channels $a$ and $b$. In an idealized scenario with identical noise budgets and equal arm lengths, the imaginary parts of the CSDs vanish, leading to identical off-diagonal terms and a symmetric covariance matrix. Under these conditions, the system can be diagonalized to obtain a set of orthogonal observables (A,E,T) via the linear transformation \cite{Prince:2002hp,Vallisneri:2007xa}
\begin{equation} \label{eq:abc2AET}
\begin{bmatrix}
\mathrm{A}_a  \\ \mathrm{E}_a  \\ \mathrm{T}_a
\end{bmatrix}
 =
\begin{bmatrix}
-\frac{1}{\sqrt{2}} & 0 & \frac{1}{\sqrt{2}} \\
\frac{1}{\sqrt{6}} & -\frac{2}{\sqrt{6}} & \frac{1}{\sqrt{6}} \\
\frac{1}{\sqrt{3}} & \frac{1}{\sqrt{3}} & \frac{1}{\sqrt{3}}
\end{bmatrix}
\begin{bmatrix}
a \\ b  \\ c
\end{bmatrix}.
\end{equation}
Although realistic arm-length mismatches can introduce non-zero imaginary components in the CSDs, the real parts dominate by several orders of magnitude for both Michelson and PD4L configurations. Consequently, the orthogonal transformation remains a valid approximation in practical applications \cite{Wang:2025mee}.

\subsection{Correlation between coefficients}

To estimate the TTL coupling coefficients $\vec{\Theta}$, we adopt the log-likelihood formulation \cite{Romano:2016dpx},
\begin{equation} \label{eq:likelihood}
\ln \mathcal{L}(d|\vec{\Theta}) = \sum_{f_i} \bigg[ -\frac{1}{2} \tilde{\mathbf{d}}_\mathrm{cor}^T \mathbf{C}^{-1} \tilde{\mathbf{d}}_\mathrm{cor} ^* - \frac{1}{2} \ln \big(\det (2\pi \mathbf{C}) \big) \bigg],
\end{equation}
where $\tilde{\mathbf{d}}_\mathrm{cor}$ denotes the frequency-domain data after TTL noise subtraction, as described in Eqs.~\eqref{eq:s_epsilon_tau}--\eqref{eq:eta}. The data simulation and analysis utilize \textsc{SATDI} \cite{Wang:2024ssp}. The matrix $\mathbf{C}$ represents the covariance of acceleration and OMS noise for the optimal TDI channels,
\begin{equation} \label{eq:cov_mat}
\mathbf{C} = \frac{T_\mathrm{obs}}{4}
\begin{bmatrix}
S_\mathrm{A} & S_\mathrm{AE} & S_\mathrm{AT} \\
S_\mathrm{EA} & S_\mathrm{E} & S_\mathrm{ET} \\
S_\mathrm{TA} & S_\mathrm{TE} & S_\mathrm{T}
\end{bmatrix},
\end{equation}
where $T_\mathrm{obs}$ denotes the observation time. As shown in \cite{Ye2025}, even when the noise spectra in Eq.~\eqref{eq:cov_mat} are treated as unknown, the inference can still converge reliably when TTL noise modeling and subtraction are incorporated.
To systematically investigate the impact of parameter degeneracy, we consider three parameter sets for the TTL coupling coefficients:
\begin{itemize}
    \item[$\vec{\Theta}_0$:] the original set of coupling coefficients associated with each angular degree of freedom;
    \item[$\vec{\Theta}_1$:] a transformed parameter set constructed from the linear (first-order) TTL coefficients,
    \begin{align}
        \mathrm{Sp}_{ij} = \mathrm{Tp}_{ij} + \mathrm{Rp}_{ij}, \label{eq:Theta1_1} \\
        \mathrm{Dp}_{ij} = \mathrm{Tp}_{ij} - \mathrm{Rp}_{ij}, \\
        \mathrm{Sy}_{ij} = \mathrm{Ty}_{ij} + \mathrm{Ry}_{ij}, \\
        \mathrm{Dy}_{ij} = \mathrm{Ty}_{ij} - \mathrm{Ry}_{ij}. \label{eq:Theta1_4}
    \end{align}
    \item[$\vec{\Theta}_2$:] a further parameter set obtained by combining coefficients in $\vec{\Theta}_1$,
    \begin{align}
        \mathrm{SDp}_i = \frac{1}{2} (\mathrm{Dp}_{ij} + \mathrm{Dp}_{ik}), \label{eq:Theta2_1} \\
        \mathrm{DDp}_i = \frac{1}{2} (\mathrm{Dp}_{ij} - \mathrm{Dp}_{ik}), \\
        \mathrm{SDy}_i = \frac{1}{2} (\mathrm{Dy}_{ij} + \mathrm{Dy}_{ik}), \\
        \mathrm{DDy}_i = \frac{1}{2} (\mathrm{Dy}_{ij} - \mathrm{Dy}_{ik}). \label{eq:Theta2_4}
    \end{align}
\end{itemize}

The motivation for these transformations is to mitigate degeneracies between TTL noise components that arise naturally during the TDI combination. As an illustrative example, the TTL noise contribution in the link from spacecraft $j$ to spacecraft $i$ is given by
\begin{equation}
\begin{aligned}
    s^\mathrm{TTL}_{ij} (t) = & \left[ \mathrm{Tp}_{ji} \phi_{ji} (t- L_{ji}) + \mathrm{Ty}_{ji} \eta_{ji} (t- L_{ji}) \right] \\
    &- \left[ \mathrm{Rp}_{ij} \phi_{ij} (t) + \mathrm{Ry}_{ij} \eta_{ij} (t) \right],
\end{aligned}
\end{equation}
while the corresponding contribution from the reverse link is
\begin{equation}
\begin{aligned}
    s^\mathrm{TTL}_{ji} (t-L_{ij}) = &  \mathrm{Tp}_{ij} \phi_{ij} (t- L_{ji} - L_{ij}) \\ 
    & + \mathrm{Ty}_{ij} \eta_{ij} (t- L_{ji} - L_{ij}) \\
    & - \mathrm{Rp}_{ji} \phi_{ji} (t - L_{ij}) \\
    & - \mathrm{Ry}_{ji} \eta_{ji} (t - L_{ij}).
\end{aligned}
\end{equation}
When these bidirectional links are combined in a TDI observable, the resulting TTL contribution becomes
\begin{equation} \label{eq:TTL_bi_links}
\begin{aligned}
    s^\mathrm{TTL}_{ij} (t) +s^\mathrm{TTL}_{ji} (t-L_{ij}) = & \mathrm{Tp}_{ij} \phi_{ij} (t- L_{ji} - L_{ij}) \\ 
    & + \mathrm{Ty}_{ij} \eta_{ij} (t- L_{ji} - L_{ij}) \\
    & + \left( \mathrm{Tp}_{ji} - \mathrm{Rp}_{ji} \right) \phi_{ji} (t - L_{ij}) \\
    & + \left( \mathrm{Ty}_{ji} - \mathrm{Ry}_{ji} \right) \eta_{ji} (t - L_{ij}) \\
    & - \left[ \mathrm{Rp}_{ij} \phi_{ij} (t) + \mathrm{Ry}_{ij} \eta_{ij} (t) \right].
\end{aligned}
\end{equation}
Eq.~\eqref{eq:TTL_bi_links} shows that the transmitting and receiving coefficient pairs $(\mathrm{Tp}_{ji},\mathrm{Rp}_{ji})$ and $(\mathrm{Ty}_{ji},\mathrm{Ry}_{ji})$ are intrinsically coupled in the TDI combination. Here, ``T'' and ``R'' denote transmitting and receiving contributions, respectively, while ``p'' and ``y'' represent pitch and yaw angular components.

\begin{figure*}[htb]
\includegraphics[width=0.98\textwidth]{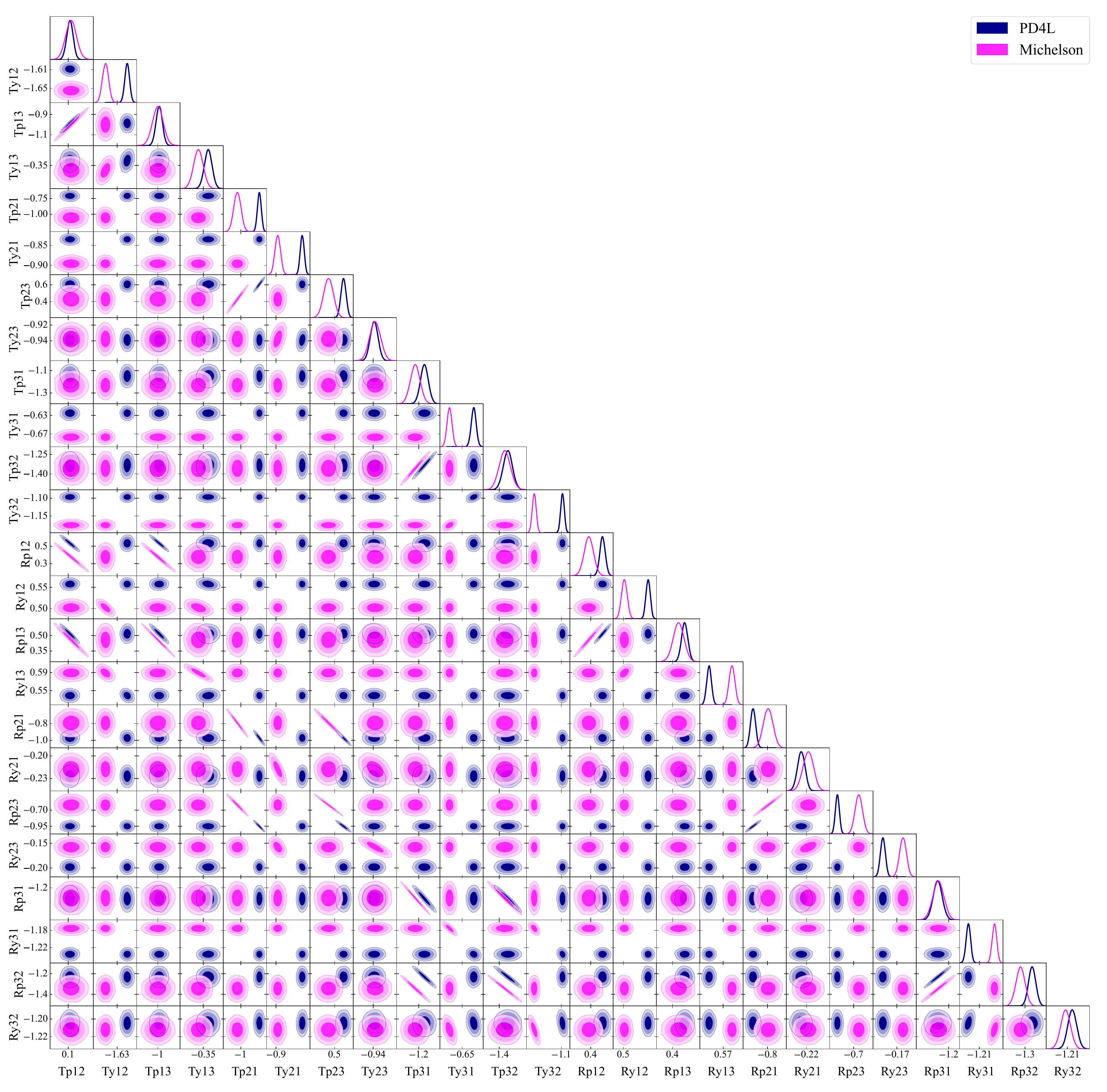}
\caption{\label{fig:posterior_invidual_parameter} Posterior distribution of the original TTL coefficients $\vec{\Theta}_0$ derived from second-generation TDI Michelson (magenta) and PD4L (blue) configurations. "T" and "R" denote the transmitter and receiver, while "p" and "y" represent the pitch and yaw components, respectively. The 1$\sigma$, 2$\sigma$, and 3$\sigma$ confidence intervals are depicted by shaded gradients, where darker hues correspond to regions of higher probability density.
}
\end{figure*}

To characterize these degeneracies, we first estimate the TTL coefficients using the linear TTL model and a Bayesian inference framework based on the nested sampler \textsc{MultiNest} \cite{Feroz:2008xx,Buchner:2014nha}. The duration of the data is set to be 7 hours, and the frequency band of [0.2 mHz, 0.1 Hz] is chosen for inference. Using the original parameter set $\vec{\Theta}_0$ with uniform priors, the resulting posterior distributions are shown in Fig.~\ref{fig:posterior_invidual_parameter}. Results obtained with the second-generation Michelson configuration are shown in magenta, while those from the PD4L configuration are shown in blue. In addition to the coupling described in Eq.~\eqref{eq:TTL_bi_links}, further degeneracies are observed among coefficient pairs such as $(\mathrm{Tp}_{ij}, \mathrm{Tp}_{ik})$, $(\mathrm{Rp}_{ij}, \mathrm{Rp}_{ik})$, and $(\mathrm{Tp}_{ij}, \mathrm{Rp}_{ik})$, consistent with previous findings \cite{Paczkowski:2022nrt}. Compared with the Michelson configuration, PD4L exhibits weaker parameter degeneracies, for example between $\mathrm{Ty}_{ij}$ coefficients.

\begin{figure*}[htb]
\includegraphics[width=0.98\textwidth]{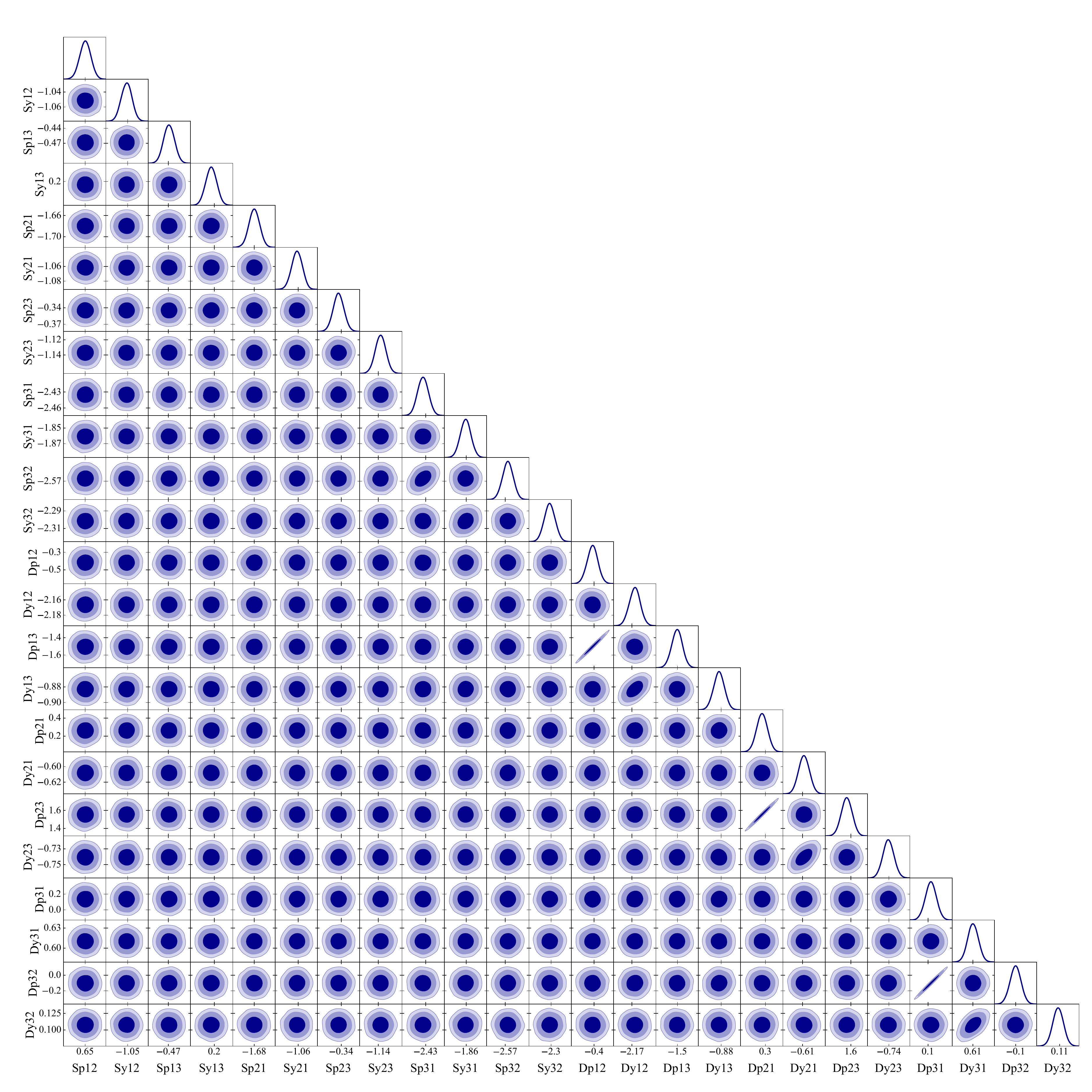}
\caption{\label{fig:posterior_combined_parameter} Posterior distribution of the TTL coefficients $\vec{\Theta}_1$ derived from PD4L TDI configuration. The parameter transformation is based on Eqs. \eqref{eq:Theta1_1}-\eqref{eq:Theta1_4} from $\vec{\Theta}_0$.
}
\end{figure*}

\begin{figure*}[htb]
\includegraphics[width=0.98\textwidth]{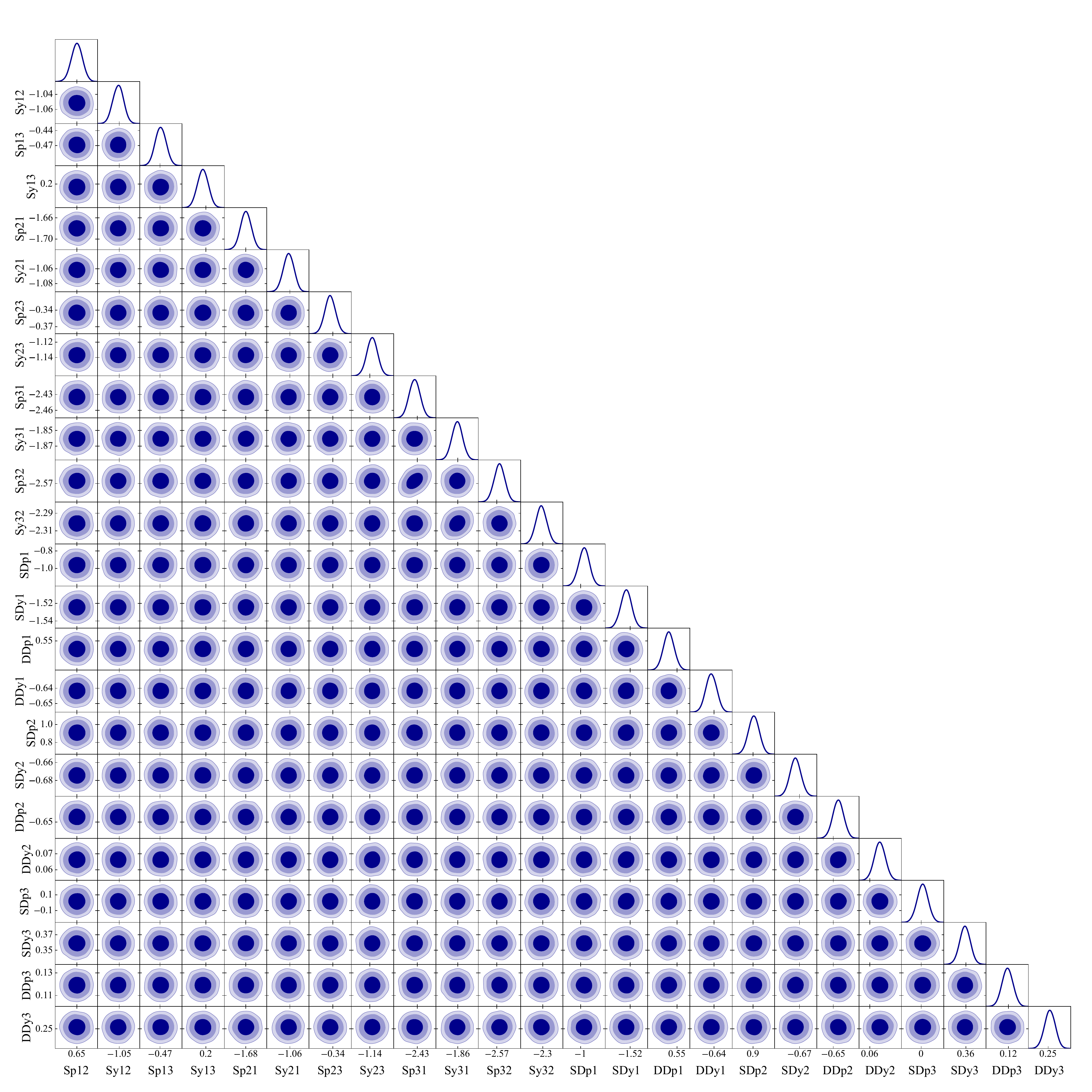}
\caption{\label{fig:posterior_combined_parameter_v2} Posterior distribution of the TTL coefficients $\vec{\Theta}_2$ obtained from PD4L TDI configuration. The parameter transformation is based on Eqs. \eqref{eq:Theta2_1}-\eqref{eq:Theta2_4} from $\vec{\Theta}_1$.
}
\end{figure*}

Applying the transformation defined in Eqs.~\eqref{eq:Theta1_1}--\eqref{eq:Theta1_4}, the posterior distributions of the transformed parameter set $\vec{\Theta}_1$ obtained with PD4L are shown in Fig.~\ref{fig:posterior_combined_parameter}. Most of the strong degeneracies present in Fig.~\ref{fig:posterior_invidual_parameter} are significantly reduced. However, several coefficient pairs, such as $(\mathrm{Dp}_{ij}, \mathrm{Dp}_{ik})$, remain strongly correlated, while moderate correlations persist for $(\mathrm{Dy}_{ij}, \mathrm{Dy}_{ik})$. To further suppress these correlations, we introduce the parameter set $\vec{\Theta}_2$ via the transformations in Eqs.~\eqref{eq:Theta2_1}--\eqref{eq:Theta2_4}. The corresponding posterior distributions are shown in Fig.~\ref{fig:posterior_combined_parameter_v2}, where the remaining degeneracies are substantially alleviated, enabling more efficient parameter inference.

\subsection{Efficient fitting}

Bayesian inference provides a comprehensive characterization of parameter uncertainties but is computationally expensive, especially for high-dimensional parameter spaces. The posterior distributions shown in Figs.~\ref{fig:posterior_invidual_parameter}--\ref{fig:posterior_combined_parameter_v2} are obtained using the linear TTL model, which involves 24 coupling coefficients. In practice, a quadratic (second-order) TTL model may be required to achieve more accurate noise modeling and subtraction, increasing the number of parameters to as many as 60. In such cases, the computational cost of Bayesian sampling becomes expensive, rendering this approach unsuitable for routine TTL noise processing.
In contrast, TTL noise subtraction requires only the most probable values of the coupling coefficients to reconstruct and remove the noise contribution. Since TTL noise arises causally from measured angular jitter through a deterministic coupling model, this structure enables stable point estimation of the coefficients using deterministic optimization methods. This observation motivates the adoption of efficient fitting algorithms.

With the three parameter sets $\vec{\Theta}_0$, $\vec{\Theta}_1$, and $\vec{\Theta}_2$, we directly estimate the TTL coupling coefficients by minimizing the likelihood function in Eq.~\eqref{eq:likelihood}. When the quadratic model is adopted, the second-order coefficients are appended to each parameter set without further transformation. All fittings are carried out using the \textsf{Minuit} package \cite{James:1975dr,iminuit}.
The computational cost under two different TTL noise levels—nominal coupling (amplification factor of 1) and increase level (amplification factor of 10)—is summarized in Table~\ref{tab:time_c} for both the Michelson and PD4L TDI configurations.

For the Michelson combination at the nominal noise level, the fitting time for the linear TTL model remains at $\sim2\times10^3$--$3\times10^3$~s across all three parameter sets, indicating that parameter redefinitions alone do not significantly reduce the intrinsic degeneracy of the Michelson response at first order. When the quadratic model is applied, the computational cost increases drastically for $\vec{\Theta}_0$ and $\vec{\Theta}_1$, reaching $10^4$~s. In contrast, the novel parameter set $\vec{\Theta}_2$ exhibits a striking reduction in fitting time to $\sim$650~s, suggesting that the transformed parameters effectively alleviate degeneracy when higher-order terms are included.

A similar but more pronounced trend is observed when the TTL noise amplitude is amplified by a factor of 10. Under this stronger coupling condition, the Michelson fitting with $\vec{\Theta}_0$ and $\vec{\Theta}_1$ still requires nearly $\sim10^4$~s in the quadratic case. However, the fitting time for $\vec{\Theta}_2$ drops to $\sim1.7\times10^3$~s, which remains substantially lower than that of the original parameterizations. Notably, in the linear model with amplified noise, the fitting time for $\vec{\Theta}_2$ decreases anomalously to 120~s, this could due to the convergent strategy of fitting algorithm for a high noise level in Michelson configuration.

\begin{table}[htbp]
    \centering
    \renewcommand{\arraystretch}{1.3}
    \setlength{\tabcolsep}{2pt}
    \caption{Computational time required for TTL coefficient fitting (in seconds). The analysis uses 7~hours of simulated data over the frequency band $[0.2\,\mathrm{mHz},\,0.1\,\mathrm{Hz}]$. Two TTL noise levels are considered: the nominal case (amplification factor of 1) and an amplified case with a factor of 10. All fits are performed on a single CPU core of an AMD EPYC~9755 processor.}
    \label{tab:time_c}
    \begin{tabular}{cc|cc|cc}
        \toprule
        & & \multicolumn{2}{c}{Michelson} & \multicolumn{2}{c}{PD4L} \\ 
        \hline
        amplification & parameter & linear & quadratic & linear & quadratic \\
        factor & set & model & model & model & model \\
        \hline
        \multirow{3}{*}{1} 
        & $\vec{\Theta}_0$ & $\sim$2700 & $\sim$12000 & $\sim$450 & $\sim$900 \\
        & $\vec{\Theta}_1$ & $\sim$2750 & $\sim$11000 & $\sim$450 & $\sim$1050 \\
        & $\vec{\Theta}_2$ & $\sim$2200 & $\sim$650   & $\sim$250 & $\sim$650 \\
        \hline 
        \multirow{3}{*}{10} 
        & $\vec{\Theta}_0$ & $\sim$2400 & $\sim$11600 & $\sim$400    & $\sim$1160 \\
        & $\vec{\Theta}_1$ & $\sim$2600 & $\sim$9900   & $\sim$430    & $\sim$2800 \\
        & $\vec{\Theta}_2$ & $\sim$120  & $\sim$1700  & $\sim$250 & $\sim$700 \\
        \hline \hline
    \end{tabular}
\end{table}

The fitting times for the alternative TDI configuration PD4L are shown in the fifth and sixth columns of Table~\ref{tab:time_c}. For both noise levels, PD4L consistently demonstrates superior computational efficiency compared to the Michelson combination. At the nominal noise level, PD4L is typically $\sim$5--6 times faster for the linear model (e.g., $\sim$450~s versus $\sim$2700~s for $\vec{\Theta}_0$), and more than an order of magnitude faster for the quadratic model in the case of $\vec{\Theta}_0$.
When the TTL noise is amplified by a factor of 10, the PD4L fitting time increases moderately but remains well controlled. For all three parameter sets, the transition from linear to quadratic modeling results in an increase in computational cost by a factor of $\sim$2--3, indicating stable and predictable scaling behavior. In contrast, the Michelson combination exhibits a much steeper and less uniform increase in fitting time, particularly for $\vec{\Theta}_0$ and $\vec{\Theta}_1$.

Overall, combining the transformed parameter set $\vec{\Theta}_2$ with the PD4L TDI configuration yields the most robust and efficient performance across both TTL noise levels. Relative to the baseline case of the Michelson configuration with the original parameters $\vec{\Theta}_0$, this approach accelerates the fitting procedure by a factor of $\sim$10 for linear modeling and up to $\sim$18 for quadratic modeling, even in the presence of strongly amplified TTL noise. This improvement is critical for enabling efficient and scalable TTL noise mitigation in realistic space-based GW data analysis pipelines.

\begin{figure*}[htb]
\includegraphics[width=0.48\textwidth]{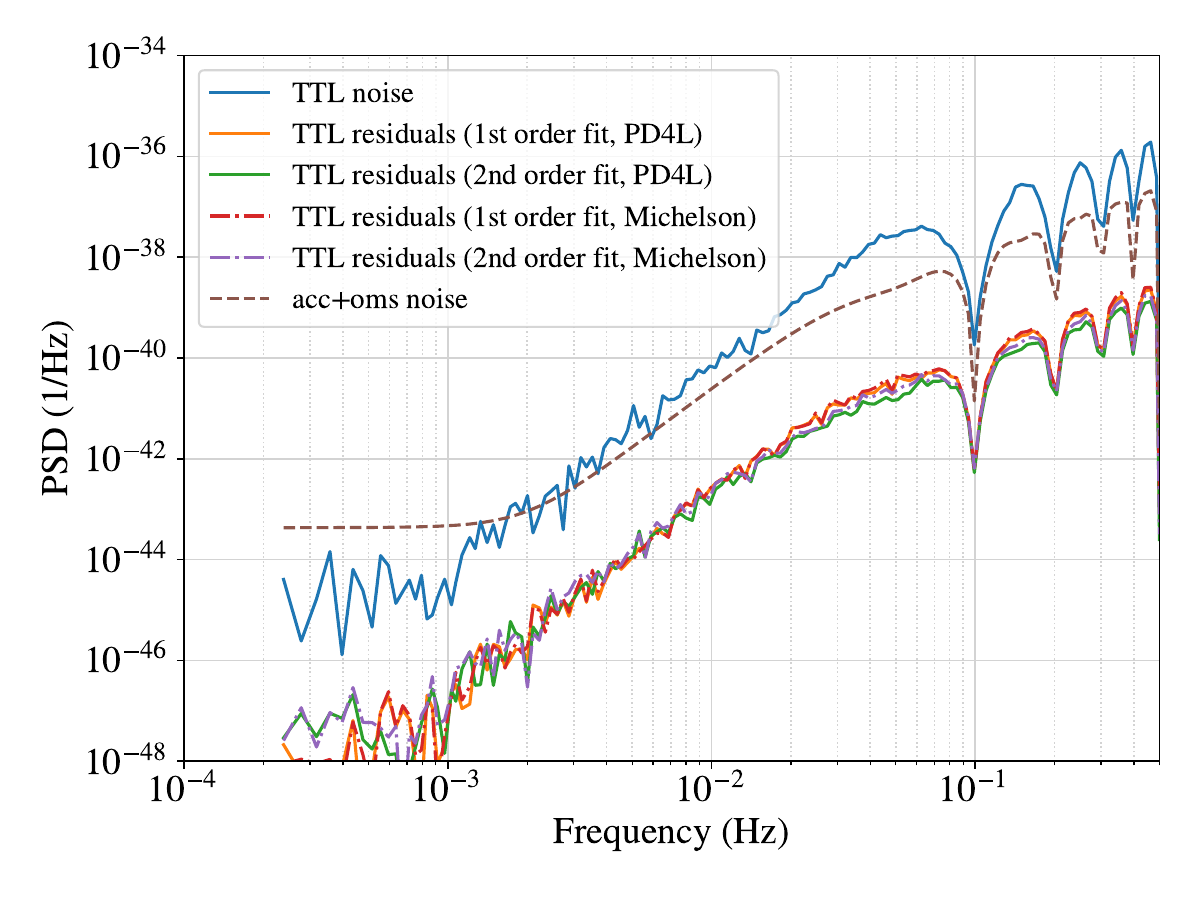}
\includegraphics[width=0.48\textwidth]{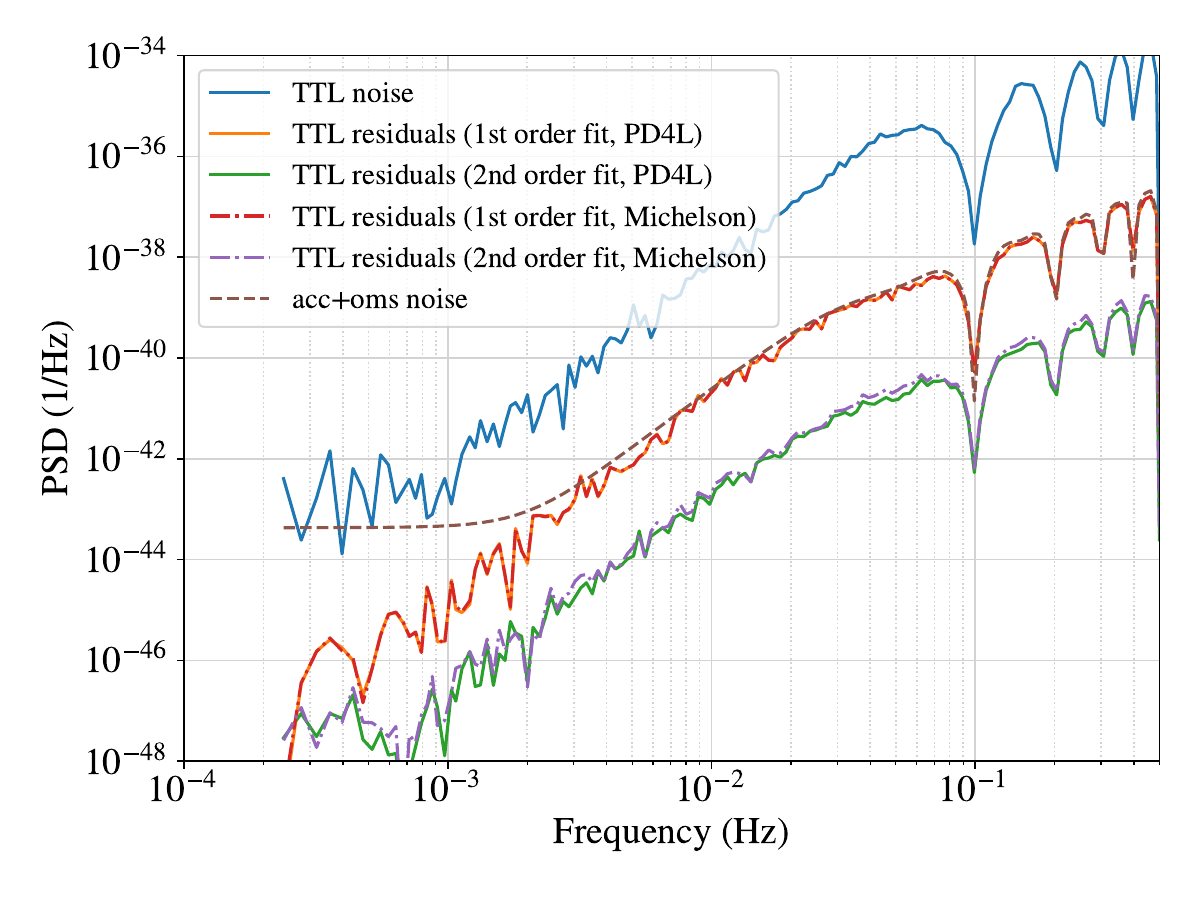}
\caption{\label{fig:TTL_subtraction_7hr}
PSDs of TTL noise and post-subtraction residuals obtained with linear and quadratic modeling. Solid blue curves show the injected TTL noise, while solid orange and solid green curves denote the residuals after subtraction using the linear (first-order) and quadratic (second-order) models using PD4L configuration, respectively. Results obtained with Michelson TDI scheme are shown for comparison, as indicated in the legend. The dashed red curves represent the reference noise floor composed of acceleration and OMS noise. 
The left panel corresponds to the nominal TTL coupling strength, whereas in the right panel the injected TTL noise amplitude is artificially amplified by a factor of 10. The quadratic model consistently achieves lower residual levels than the linear model, particularly in the amplified case, where it remains below the acc+oms noise floor across the full frequency band.
}
\end{figure*}

Fig.~\ref{fig:TTL_subtraction_7hr} illustrates the performance of the TTL noise subtraction in the frequency domain using the $\Theta_2$ fitting. The PSDs are shown for the injected TTL noise (solid blue), by using the PD4L configuration, the residuals after subtraction using the linear model (solid orange), the residuals obtained with the quadratic model (solid green), and the reference noise floor (dashed red), which consists of acceleration and OMS noise.
In the nominal coupling scenario (left panel), the injected TTL noise dominates the spectrum above 2 mHz. Applying the linear model significantly suppresses the noise, reducing the residual level by approximately two orders of magnitude and bringing it close to the acc+oms (acceleration and OMS) noise floor. The quadratic model provides a further, though slightly, improvement, yielding residuals that remain well below the reference noise level across the entire frequency band of interest.

The advantage of the higher-order model becomes more pronounced when the TTL coupling strength is increased. The right panel of Fig.~\ref{fig:TTL_subtraction_7hr} shows the case in which the injected TTL noise amplitude is amplified by a factor of 10. Under this more demanding condition, the linear model becomes inadequate: its residuals (orange curves) approach the acc+oms noise floor at frequencies higher than 2~mHz band, leaving a non-negligible level of unmitigated noise. In contrast, the quadratic model (green curves) demonstrates substantially improved robustness, successfully suppressing the residuals below the acc+oms limit even in the presence of strongly amplified TTL noise.
This comparison indicates that while a linear approximation is sufficient for weak TTL coupling, the inclusion of second-order non-linear terms is essential for maintaining the required sensitivity when the TTL noise level is high.
Compared with the results obtained using the Michelson configuration, shown by the dash-dotted curves, the residual noise levels are found to be comparable to those achieved with PD4L. However, as summarized in Table~\ref{tab:time_c}, fitting the coupling coefficients using the PD4L configuration together with the $\Theta_2$ parameter set provides a more robust and computationally efficient approach.

\section{Conclusions and Discussion} \label{sec:conclusions}

In this work, we have investigated the post-processing calibration and subtraction of TTL noise in space-based GW detectors using second-generation TDI observables. By explicitly modeling the coupling between angular jitter and optical path length fluctuations, we examined both linear (first-order) and quadratic (second-order) TTL noise models under realistic instrumental noise conditions.
A key result of this study is the identification and mitigation of strong parameter degeneracies inherent in the conventional TTL coefficient representation. Through a sequence of linear transformations, we constructed alternative parameter set that progressively reduce the coupling between coefficients. These transformations significantly simplify the effective parameter space explored during fitting, without altering the physical content of the TTL model.

We demonstrated that the choice of TDI configuration plays a critical role in the efficiency of TTL parameter estimation. Compared to the fiducial second-generation Michelson combination, the alternative PD4L configuration exhibits substantially weaker parameter degeneracies and markedly improved numerical conditioning. As a result, the direct likelihood-based fitting converge more rapidly when PD4L observables are employed. When combined with the most effective parameter set, this improvement leads to an overall acceleration of the fitting procedure by approximately one order of magnitude for the linear model and up to $\sim$18 for the quadratic model.

Our results further clarify the regimes in which higher-order TTL modeling becomes necessary. For nominal coupling strengths, the linear model is sufficient to suppress TTL noise well below the acceleration and OMS noise floor. However, when the TTL coupling amplitude is increased, the residuals from the linear model become comparable to the instrumental noise at high frequencies, potentially limiting the detector sensitivity. In contrast, the quadratic model consistently achieves robust suppression across the full frequency band, even in the presence of strongly amplified TTL noise. This behavior highlights the importance of nonlinear modeling for future missions operating with tighter sensitivity requirements or increased angular jitter.

From a methodological perspective, the reduced parameter degeneracy enables the replacement of computationally intensive Bayesian samplers with efficient deterministic optimization algorithms such as \textsf{Minuit}. Since TTL subtraction requires only the maximum-likelihood estimates of the coupling coefficients, rather than their full posterior distributions, this approach is particularly well suited for large-scale data pre-analysis pipelines and iterative calibration schemes.

Several extensions of this work merit further investigation. The present analysis assumes stationary noise spectra in the construction of optimal TDI channels, and the jitter angles are accurately measured without drift in seven hours. Incorporating time-dependent noise characteristics, measurement errors, and correlations with other instrumental systematics may further refine the robustness of the subtraction. In addition, applying the proposed parameterization and fitting strategy to longer data segments and to full mission simulations will be essential for assessing its performance in realistic end-to-end analyses.

In summary, we have shown that a judicious combination of alternative TDI observables and optimized TTL parameterization can dramatically improve the efficiency and robustness of TTL noise calibration. These advances provide a practical pathway toward scalable and accurate noise mitigation for future space-based GW detectors.

This work are performed by using the python packages \textsf{numpy} \cite{harris2020array}, \textsf{scipy} \cite{2020SciPy-NMeth}, \textsf{pandas} \cite{pandas}, \textsf{MultiNest} \cite{Feroz:2008xx} and \textsf{PyMultiNest} \cite{Buchner:2014nha}, \textsf{Minuit} \cite{James:1975dr,iminuit}, and the plots are make by utilizing \textsf{matplotlib} \cite{Hunter:2007ouj}, \textsf{GetDist} \cite{Lewis:2019xzd}. 

\begin{acknowledgments}
G.W. was supported by the National Key Research and Development Program of China under Grant No. 2021YFC2201903 and NSFC Grant No. 12575058.
X.W. was supported by the National Key Research and Development Program of China under Grant  No.2024YFC2206900.
\end{acknowledgments}

\appendix

\nocite{*}

\bibliography{ref}

\end{document}